\documentclass[preprint]{aastex}
\usepackage{color}

\newcommand{\ME}{M_{\rm E}}
\newcommand{\sub}[1]{_{\rm #1}}
\newcommand{\super}[1]{^{\rm #1}}

\slugcomment{ApJ, in press}

\shorttitle{CONSTRAINTS ON THE MASS OF A HABITABLE PLANET}
\shortauthors{Ikoma \& Genda}

\begin{document}
%\setlength{\baselineskip}{20pt}


\title{CONSTRAINTS ON THE MASS OF A HABITABLE PLANET WITH WATER OF NEBULAR ORIGIN}


\author{Masahiro Ikoma\altaffilmark{}}
\affil{Research Center for the Evolving Earth and Planets, 
Tokyo Institute of Technology, Ookayama, Meguro-ku, Tokyo 152-8551, 
Japan}
\email{mikoma@geo.titech.ac.jp}

\and

\author{Hidenori Genda\altaffilmark{}}
\affil{Department of Earth and Planetary Sciences, 
Tokyo Institute of Technology, Ookayama, Meguro-ku, Tokyo 152-8551, 
Japan}
\email{genda@geo.titech.ac.jp}


%\altaffiltext{2}{Society of Fellows, Harvard University.}


\begin{abstract}
From an astrobiological point of view, 
special attention has been paid to the probability of 
habitable planets in extrasolar systems. 
The purpose of this study is to constrain a possible range of 
the mass of a terrestrial planet that can get water. 
We focus on the process of water production through 
oxidation of the atmospheric hydrogen---the nebular gas 
having been attracted gravitationally---by oxide 
available at the planetary surface. 
For the water production to work well on a planet, 
a sufficient amount of hydrogen and enough high temperature 
to melt the planetary surface are needed. 
We have simulated the structure of the atmosphere 
that connects with the protoplanetary nebula 
for wide ranges of heat flux, opacity, and density of the nebular gas.  
We have found both requirements are fulfilled 
for an Earth-mass planet for wide ranges of the parameters. 
We have also found the surface temperature of planets of $\leq 0.3 \ME$ 
($\ME$: Earth's mass) is lower than the melting temperature of silicate 
($\sim 1500$~ K). 
On the other hand, a planet of more than several~$\ME$ becomes 
a gas giant planet through runaway accretion of the nebular gas. 
\end{abstract}


%% subject header.  Objects should be in the appropriate "individual"
%% headers (e.g. quasars: individual, stars: individual, etc.) with the
%% additional provision that the total number of headers, including each
%% individual object, not exceed six.  The \objectname{} macro, and its
%% alias \object{}, is used to mark each object.  The macro takes the object
%% name as its primary argument.  This name will appear in the paper
%% and serve as the link's anchor in the electronic edition if the name
%% is recognized by the data centers.  The macro also takes an optional
%% argument in parentheses in cases where the data center identification
%% differs from what is to be printed in the paper.



\section{INTRODUCTION}

More than 170 extrasolar planets have been detected so far; 
most of them are Jupiter-like planets (see www.obspm.fr/planets). 
At present, several projects are in progress 
to discover terrestrial planets (e.g., TPF, Darwin, etc.). 
Special attention has thus been paid to habitable planets 
in extrasolar systems. 
A prerequisite for a planet being habitable is considered 
to be the existence of liquid water on it. 
For a planet to retain liquid water on its surface, 
it must acquire a sufficient amount of water and be located 
at a suitable distance from its parent star. 
This paper focuses on the former issue and 
constrains the probability of the existence of habitable planets 
in extrasolar systems. 

The latter issue has been discussed using the term ``habitable zone''. 
The habitable zone (HZ) is defined as a range of orbital distance 
from a star within which a planet can retain liquid water on its surface. 
Around a solar-mass main-sequence star, 
the HZ is located around 1~AU and its width is as narrow as $\sim$0.4~AU 
\citep{KWR93}.
The surface temperature of a planet beyond the HZ 
is below the freezing temperature of $\rm H_2O$, 
so that the planet is unable to keep liquid water continuously. 
Because of high surface temperature of a planet 
closer to its parent star than the HZ, 
the concentration of $\rm H_2O$ in the upper atmosphere 
is so high that the planet suffers substantial escape of $\rm H_2O$ 
due to incident stellar UV radiation, 
and ends up losing its ocean completely. 

Formation of terrestrial planets in that narrow zone, however, 
seems to be by no means unlikely within the context of the core accretion model 
for planet formation. 
\citet{IL04a,IL04b,IL05}, for example, constructed an integrated model 
for planet formation including the accretion and dynamical evolution of planets. 
Their model reproduced the mass-period distribution of detected extrasolar planets 
and also predicted that of unknown planets. 
In the predicted mass-period distribution, 
the HZ is filled with hypothetical terrestrial planets 
of various masses \citep[e.g., Figure~12 of][]{IL04a}. 
This means the existence of planets in the HZ is likely; 
the remaining issue is thus
how likely a planet acquires a sufficient amount of water. 

In this paper, we consider the nebular origin of water of terrestrial planets. 
A planet embedded in a protoplanetary nebula attracts gravitationally 
the nebular gas to have a hydrogen-rich atmosphere \citep*{HNM79,MNH78}. 
The atmospheric hydrogen can be oxidized by some oxide 
contained in the planet to produce water on the planet, 
which was first proposed by \citet{S90}. 

The reason why we focus on the process of water production is that 
it could commonly happen on any extrasolar terrestrial planet. 
Planets are in general formed in hydrogen-rich protoplanetary nebulae. 
Oxides are also available on a terrestrial planet 
if the C/O ratio of the system is less than unity \citep{L75,LB79}; 
most main-sequence stars are known to have C/O ratios less than unity 
\citep[e.g.,][]{RTLA03,TH05}. 
The remaining requirements are a sufficient amount of hydrogen 
and enough high temperature to melt the surface of the planet; 
the molten planetary surface being called the magma ocean. 
The second condition is needed because 
if the planetary surface remains solid, 
the reaction between the atmospheric hydrogen and the surface oxide 
would result just in production of membrane covering the surface, 
not leading to production of a large amount of water. 

In this paper, we investigate properties of the atmosphere of nebular origin. 
We first clarify the conditions for a planet 
to have a massive hydrogen-rich atmosphere and a magma ocean 
to constrain the range of the mass of a planet 
that has a sufficient amount of water. 
The structure of the nebular-origin atmosphere was investigated 
by \citet{HNM79} and \citet{NMSH85} 
for wide ranges of planetary accretion rate, grain opacity, 
and density of the nebular gas. 
However, they used quite simple forms of opacity and equation of state, 
both of which have been substantially improved. 
Furthermore they focused only on the Earth (i.e., Earth-mass planets) 
and had no discussion on the production of water. 
Although \citet{S90} discussed the production of water 
in order to suggest the deep magma ocean on the early Earth, 
the ranges of the parameters considered were so restricted 
that we are unable to get any systematic understanding 
of the nebular origin of water 
on terrestrial planets. 

We also constrain the mass of a planet that remains a terrestrial one. 
Habitable planets may be not gas giant but terrestrial ones. 
If a planet is isolated from planetesimals, 
it captures a substantial amount of the nebular gas to become a gas giant planet 
\citep*{INE00}. 
We simulate the evolution and accumulation of the planetary atmosphere 
to obtain the timescale for the substantial gas accretion 
as a function of planet's mass. 
The timescale is compared to the lifetime of the nebular gas 
of $10^6$--$10^7$ years \citep*{NGM00} 
to constrain a possible range of the mass of a habitable planet. 

Section~2 describes our numerical method. 
Section~3 presents properties of the atmosphere of nebular origin 
for various planetary masses. 
Section~4 shows the timescale for the substantial accretion of the nebular gas. 
Finally we discuss the probability of water production on terrestrial planets 
and constrain the masses of the potentially-habitable planets in section~5.

\section{NUMERICAL METHOD}
\subsection{\it Basic equations}
We consider a spherically-symmetric hydrostatic atmosphere 
that connects with the surrounding nebula. 
The atmospheric structure is determined by the equation of hydrostatic equilibrium 
including the self-gravity of the atmosphere,
\begin{equation}
\frac{\partial P}{\partial r} = - \frac{G M_r \rho}{r^2},
\label{eq: hydrostatic equilibrium}
\end{equation}
and the equation of mass conservation,
\begin{equation}
\frac{\partial M_r}{\partial r} = 4 \pi r^2 \rho,
\label{eq: mass conservation}
\end{equation}
where $P$ and $\rho$ are respectively pressure and density 
of atmospheric gas, $r$ is distance from the planet's center, 
$M_r$ is mass inside a sphere of radius $r$, 
and $G$ is the gravitational constant. 

The thermal structure is determined by energy transport. 
If the optical thickness that is defined by 
\begin{equation}
\tau \equiv \int_r^{R} \kappa \rho dr
\label{eq: optical depth}
\end{equation}
is smaller than $2/3$, temperature is given by \citep{HNM79}
\begin{equation}
T^4 = T^4\sub{n} + 
\frac{L}{8 \pi \sigma\sub{SB} r^2}
\frac{1+3\tau/2}{2-3\tau/2};
\label{eq: isothermal}
\end{equation}
in equations~(\ref{eq: optical depth}) and (\ref{eq: isothermal}), 
$\tau$ is the optical thickness, $\kappa$ is the Rosseland mean opacity, 
$R$ is the outer radius of the atmosphere defined below, 
$T$ and $T\sub{n}$ are respectively atmospheric and nebular gas temperatures, 
and $\sigma\sub{SB}$ is the Stefan-Boltzmann constant. 
Equation~(\ref{eq: isothermal}) is similar to 
the well-known formula for radiative transfer 
in a plane-parallel gray atmosphere,  
but includes roughly the effect of spherical geometry. 
Despite of its roughness, we use this convenient formula 
because quantities of our interest (e.g., surface temperature) 
are insensitive to the structure of the optically-thin layer. 

If $\tau > 2/3$, temperature distribution is determined by 
the smaller of adiabatic and radiative temperature gradients \citep{KW94}: 
The former is given by 
\begin{equation}
\frac{\partial T}{\partial r} = 
- \frac{G M_r \rho}{r^2} \left( \frac{\partial T}{\partial P} \right)_S;
\label{eq: adiabatic temperature gradient}
\end{equation}  
the latter is given by 
\begin{equation}
\frac{\partial T}{\partial r} = 
- \frac{3 \kappa \rho L}{64 \pi \sigma\sub{SB} r^2 T^3},
\label{eq: radiative temperature gradient}
\end{equation}
where $S$ is specific entropy and 
$L$ is energy flux passing through a sphere of radius $r$ 
(called luminosity). 
We assume the opacity is the sum of the Rosseland mean opacities 
of gas molecules and dust grains (see below). 
The luminosity is determined by \citep{KW94}
\begin{equation}
\frac{\partial L}{\partial r} = - 4 \pi r^2 \rho T \frac{dS}{dt},
\label{eq: energy conservation}
\end{equation} 
where we have assumed there is no energy generation other than 
entropy change in the atmosphere. 

Two types of simulation are done in this paper, quasi-static and static simulations. 
To calculate the timescale of the gas accretion in section~4, 
we perform quasi-static simulations integrating all the above equations 
but equation~(\ref{eq: isothermal}). 
Inclusion of equation~(\ref{eq: isothermal}) makes the simulation so complicated, 
but yields little change in the results obtained in section~4. 
To investigate the properties of the atmosphere extensively in section~3, 
we perform static simulations integrating 
equations~(\ref{eq: hydrostatic equilibrium})-(\ref{eq: radiative temperature gradient}), 
because the quasi-static simulations are time-consuming. 
In the static simulations we assume $L$ is spatially constant, 
instead of solving equation~(\ref{eq: energy conservation}); 
this assumption was shown to be appropriate 
for less massive atmospheres \citep{INE00}. 

\subsection{\it Input physics}
The equation of state used here is the nonideal one given by \citet{SCVH95}, 
interpolated to a composition of $X = 0.74$, $Y = 0.24$, and $Z = 0.02$, 
where $X$, $Y$, and $Z$ are the mass fractions of hydrogen, helium, 
and elements heavier than helium, respectively. 

The grain opacity is taken from \citet*{PMC85}, 
who presumed dust grains with an interstellar size distribution. 
Since the amount and size distribution of dust grains in the atmosphere 
are highly uncertain, 
we regard the grain opacity as a parameter using the conventional form of 
$\kappa\sub{gr} = f \kappa\sub{gr}\super{P}$, 
where $\kappa\sub{gr}$ is the grain opacity, 
$\kappa\sub{gr}\super{P}$ is that given by \citet{PMC85}, 
and $f$ is the grain depletion factor. 

The gas opacity is taken mainly from \citet{AF94}, 
supplemented with the data of \citet{M80};  
compared to the latter, the opacity for $\rm H_2O$ has been substantially improved 
in the former. 
In \citet{AF94}, data are available for limited ranges of temperature and density, 
$2.8 \leq \log T \leq 4.1$ and
$-2 \leq \log Q \leq 3$, 
where $Q = \rho/T_6^3$, $\rho$ is density in $\rm g\,cm^{-3}$, 
and $T_6$ is temperature in millions of degrees. 
In the calculations here, we need data for lower temperature and higher density. 
For $\log T \geq 3.5$ both data sets are similar to each other, 
because a major source of the opacity is hydrogen for the temperature range. 
We thus use \citet{M80}'s opacity for $\log Q > 3$ and $\log T \geq 3.5$. 
For $2.8 \leq \log T \leq 3.5$, a major source of the opacity is $\rm H_2O$, 
and the opacity is rather insensitive to the density. 
We thus extrapolate values of \citet{AF94} to $\log Q > 3$ 
for that temperature range. 
For $\log T \leq 2.8$, we adopt values at $\log T = 2.8$, 
because the opacity is nearly constant for $2.8 \leq \log T \leq 3.5$. 

As mentioned in Introduction, 
previous workers used rather simple forms of opacity and equation of state. 
Hydrogen molecules dissociate to hydrogen atoms 
for the ranges of temperature and density considered here. 
\citet{HNM79}, however, used the equation of state for a perfect gas, 
although the effect of the dissociation was included 
in the calculation of the adiabatic temperature gradient. 
The form of the grain opacity adopted by \citet{HNM79} and \citet{INE00}
was quite simple; 
the value of the grain opacity was constant 
until dust grains evaporate at 1500~K independent of gas density. 
However, the Rosseland mean values of the wavelength-dependent opacity 
of dust grains depend on temperature and, in reality, 
dust grains consist of several different components 
whose evaporation temperatures depend on gas density. 
Furthermore, as mentioned above, the gas opacity of $\rm H_2O$ has been improved 
and is rather high compared to that used by those workers.
For low temperature $\log T < 3.5$, 
$\rm H_2O$ makes a dominant contribution to the gas opacity. 

\subsection{\it Boundary conditions}
Four boundary conditions are needed. 
Two of the four are given at the bottom of the atmosphere as
\begin{equation}
 \begin{array}{ccccc}
 \displaystyle{
 M_r = \frac{4 \pi r^3}{3} \rho\sub{s}
 } &
 \mbox{and} & L = L\sub{s} & \mbox{at} & 
 \displaystyle{
 r = R\sub{s} \equiv \left( \frac{3 M\sub{s}}{4 \pi \rho\sub{s}} \right)^{1/3}, }
 \label{eq: inner boundary conditions}
 \end{array}
\end{equation}
where $\rho\sub{s}$, $M\sub{s}$, and $R\sub{s}$ are 
the mean density (= 3.9~$\rm g \, cm^{-3}$), mass, and radius 
of the solid part of the planet (called simply the solid planet hereafter), 
respectively, and $L\sub{s}$ is the energy flux at the surface of the solid planet. 

The other two boundary conditions are given at the outer edge of the atmosphere. 
The outer radius is assumed to be the smaller of 
the Bondi and the Hill radii, which are respectively defined by
\begin{equation}
R\sub{B} = \left( 1-\frac{1}{\gamma} \right)\frac{G M\sub{p}}{c\sub{T}^2}
\label{eq: Bondi radius}
\end{equation}
and
\begin{equation}
R\sub{H} = \left( \frac{M\sub{p}}{3 M_\ast} \right)^{1/3} a,
\label{eq: Hill radius}
\end{equation}
where $M\sub{p}$ is planetary total mass, 
$\gamma$ and $c\sub{T}$ are respectively adiabatic exponent and 
isothermal sound speed of the nebular gas, 
$a$ is heliocentric distance, and 
$M_\ast$ is mass of the parent star. 
At the Bondi radius the enthalpy of gas is equal to 
the potential energy of gas, while 
at the Hill radius planetary gravity is equal to 
the tidal force due to the parent star. 
For an Earth-mass planet, $R\sub{B}$ and $R\sub{H}$ are approximately 
$20 R\sub{s}$ and $200 R\sub{s}$. 

At the outer boundary the atmosphere is assumed 
to connect smoothly with the nebular gas, namely,
\begin{equation}
 \begin{array}{ccccc}
 T = T\sub{n} & \mbox{and} & P = P\sub{n} & \mbox{at} & r = R,
 \end{array}
\label{eq: outer boundary conditions}
\end{equation}
where $T\sub{n}$ and $P\sub{n}$ are the temperature and pressure of the nebular gas. 
In this paper, $M_\ast$ is the solar mass, $a = 1 \rm AU$, 
and $T\sub{n} = 280 \rm K$, 
which is the temperature at 1~AU in the minimum-mass solar nebula \citep{H81}. 
The nebular pressure (equivalently the nebular density) is regarded 
as a parameter. 

\subsection{\it Parameters}
The parameters in our atmospheric model for a given planet's mass ($M\sub{s}$) 
are luminosity ($L\sub{s}$), grain depletion factor ($f$), 
and the nebular density ($\rho\sub{n}$). 

Because luminosity is supplied by incoming planetesimals 
in accretion stages, 
$L\sub{s}$ is approximately expressed by 
\begin{equation}
L\sub{s} \simeq \frac{GM\sub{s}}{R\sub{s}} \dot{M}\sub{s}
= 1.1 \times 10^{24}
\left(\frac{M\sub{s}}{1\ME}\right)^{2/3}
\left(\frac{\dot{M}\sub{s}}{10^{-8}\ME\,{\rm yr}^{-1}}\right)
\left(\frac{\rho\sub{s}}{3.9 {\rm g\,cm^{-3}}}\right)^{1/3} 
\rm erg\,s^{-1},
\label{eq: accretion luminosity}
\end{equation}
where $\dot{M}\sub{s}$ is planetesimal accretion rate. 
The planetesimal accretion rate in late stages of planet formation is uncertain. 
However, in this paper, since we consider the situations 
where formation of terrestrial planets is completed 
before the nebular gas disappears ($10^6$--$10^7$ years), 
$M\sub{s}/\dot{M}\sub{s}$ should be longer than $\sim 1\times10^8$ years 
at the time when the planet has grown in mass close to its current mass; 
otherwise, the planet appreciably grows after the disappearance of the nebula. 
If $M\sub{s}/\dot{M}\sub{s} > 1 \times 10^8$~years, 
$L\sub{s} < 1\times10^{24} \rm erg\,s^{-1}$ 
for an Earth-mass planet and $L\sub{s} < 1\times10^{23} \rm erg\,s^{-1}$ 
for a Mars-mass ($\sim$ 0.1~$\ME$) planet. 
A lower limit of luminosity might be constrained by radiogenic luminosity. 
On the early Earth and Mars, the values of radiogenic luminosity are 
$\sim 1 \times 10^{21} \rm erg\,s^{-1}$ \citep{OP98} and  
$\sim 1 \times 10^{20} \rm erg\,s^{-1}$ \citep{WD88}.

The appropriate range of $f$ is also uncertain. 
Although \citet{P03} suggested $f \sim 0.01$ through 
the numerical simulations of coagulation and sedimentation of 
dust grains in the atmosphere, 
more extensive study is needed because their simulations were performed 
only in a few atmospheric models. 
We thus consider $f = 0$--$1$ in this paper. 

The nebular gas dissipates with time. 
Taking it into account, we consider a wide range of the nebular density, 
1--$10^{-10}$ times the gas density 
($\rho\sub{MSN} = 1.2 \times 10^{-9} \rm g\,cm^{-3}$ at 1~AU)  
of minimum-mass solar nebula model \citep{H81} for the Earth-mass case. 

\section{PROPERTIES OF THE ATMOSPHERE}
\begin{figure}
\includegraphics[width=8cm,keepaspectratio]{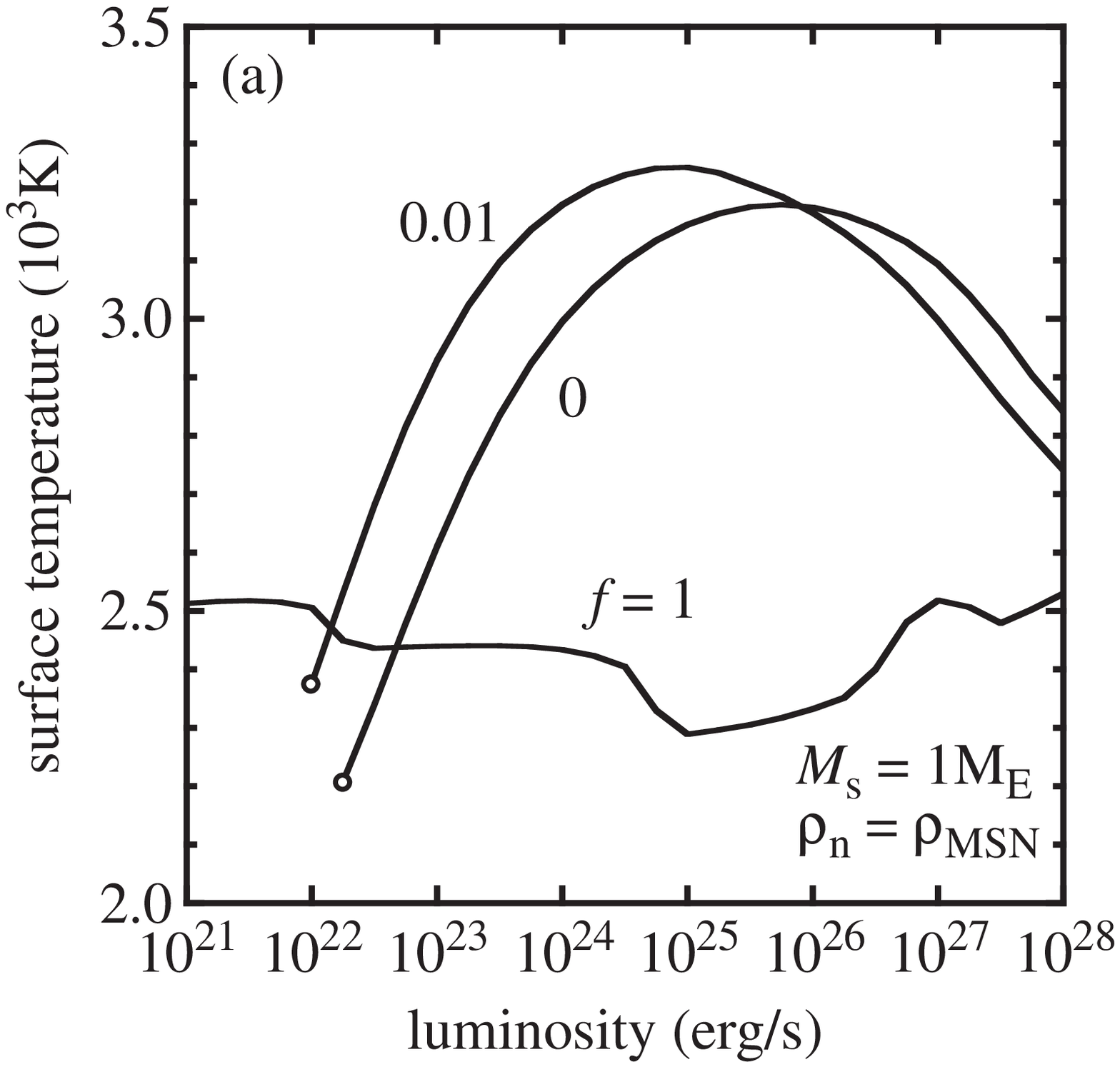}
\includegraphics[width=8cm,keepaspectratio]{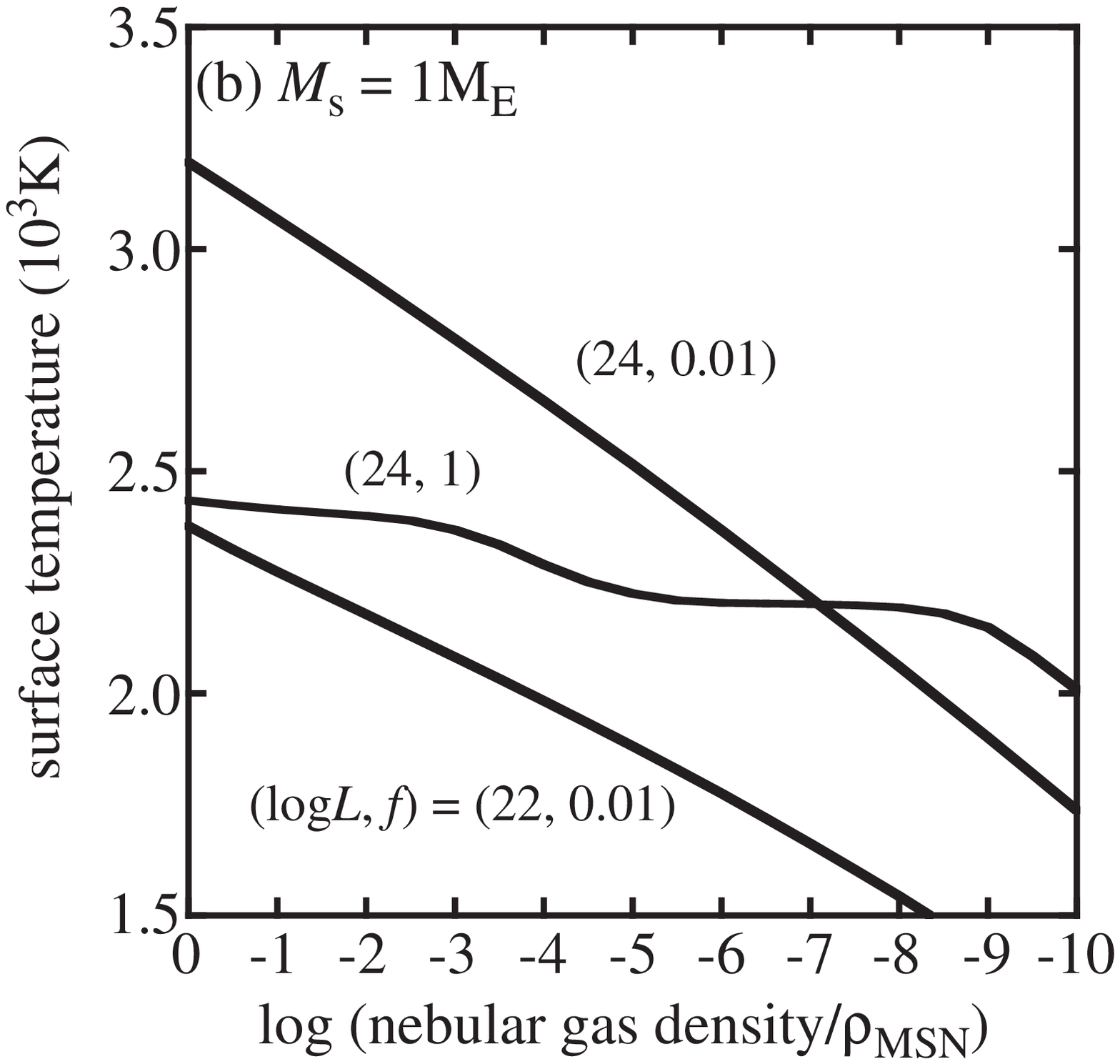}
\caption{The surface temperature (i.e., the temperature 
at the bottom of the atmosphere) of an Earth-mass planet 
for wide ranges of three parameters, luminosity ($L$), 
grain depletion factor ($f$), and density of the nebular gas ($\rho\sub{n}$).
In (a), the surface temperature is shown as a function of $L$ 
for three different values of $f$. 
In (b), the surface temperature is shown 
as a function of $\rho\sub{n}$ normalized by that in the minimum-mass solar nebula 
($\rho\sub{MSN}= 1.2 \times 10^{-9} \rm g \, cm^{-3}$) 
for three different sets of values of $L$ and $f$. 
The attached sets of values are $\log L$ in $\rm erg\,s^{-1}$ 
and $f$ in (b). 
Circles represent critical luminosities (see the text). 
}
\label{f1}
\end{figure}

Figure~\ref{f1}a illustrates that the surface temperature of an Earth-mass planet 
changes only by at most 50~\% despite of the order-of-magnitude differences 
in luminosity and grain opacity, 
and is always higher than 2000~K when $\rho\sub{n} = \rho\sub{MSN}$. 
The shapes of the functions for $f = 0$ and $0.01$ 
are convex. 
The atmosphere is largely radiative for lower luminosities, 
whereas it is largely convective for higher luminosities. 
Although surface temperature in general increases with luminosity 
for radiative atmospheres, its dependence is found to be weak. 
This is because the effect of higher luminosity is almost compensated for with 
that of smaller optical depth (i.e., smaller mass of the atmosphere) 
that is due to high luminosity as shown below (see Fig.~\ref{f3}a). 
For fully convective atmospheres, 
any atmospheric property is independent of luminosity. 
As luminosity increases, the inner convective region expands outward, 
but, at the same time, the optically-thin, almost isothermal layer 
also extends inward because of a decrease in atmospheric mass. 
The latter effect is responsible for the decrease in surface temperature 
for high luminosities. 
The complicated form of the function for $f = 1$ comes from 
the complicated form of the grain opacity; 
several different components evaporate at different temperatures 
depending on gas density. 

\begin{figure}
\includegraphics[]{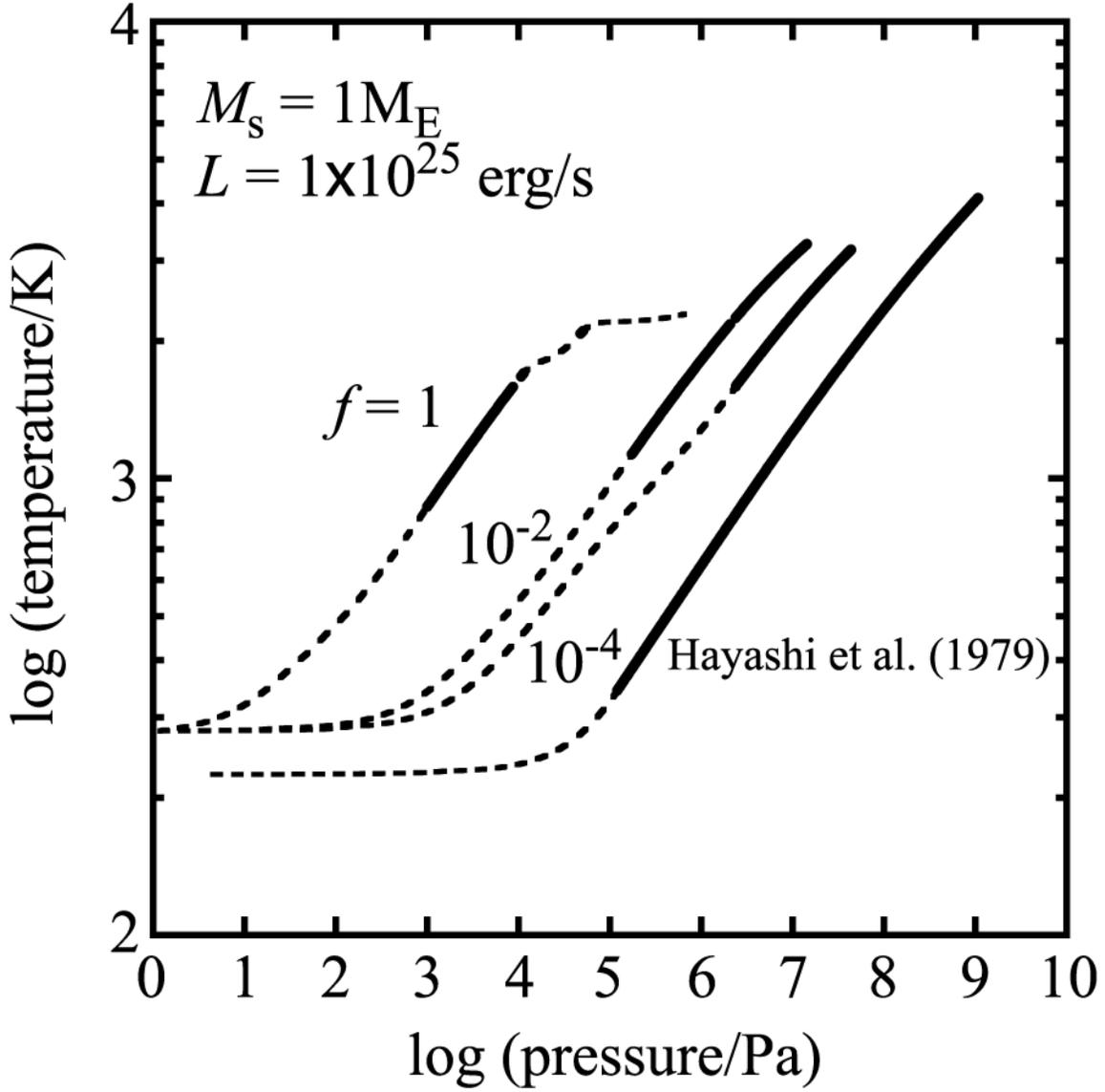}
\caption{The atmospheric structure of an Earth-mass planet 
for three different values of grain depletion factor ($f$) and 
luminosity ($L$) of $1\times10^{25} \rm erg\,s^{-1}$. 
The structure simulated in the similar way to \citet{HNM79} 
is also shown; 
in the simulation we have adopted 
$\rho\sub{n} = 5.7 \times 10^{-9} \rm g\,cm^{-3}$ 
and $T\sub{n} = 225 \rm K$, following \citet{HNM79}. 
The bold and dashed lines represent convective and radiative layers, respecitively.
}
\label{f2}
\end{figure}

\begin{figure}
\includegraphics[width=8cm,keepaspectratio]{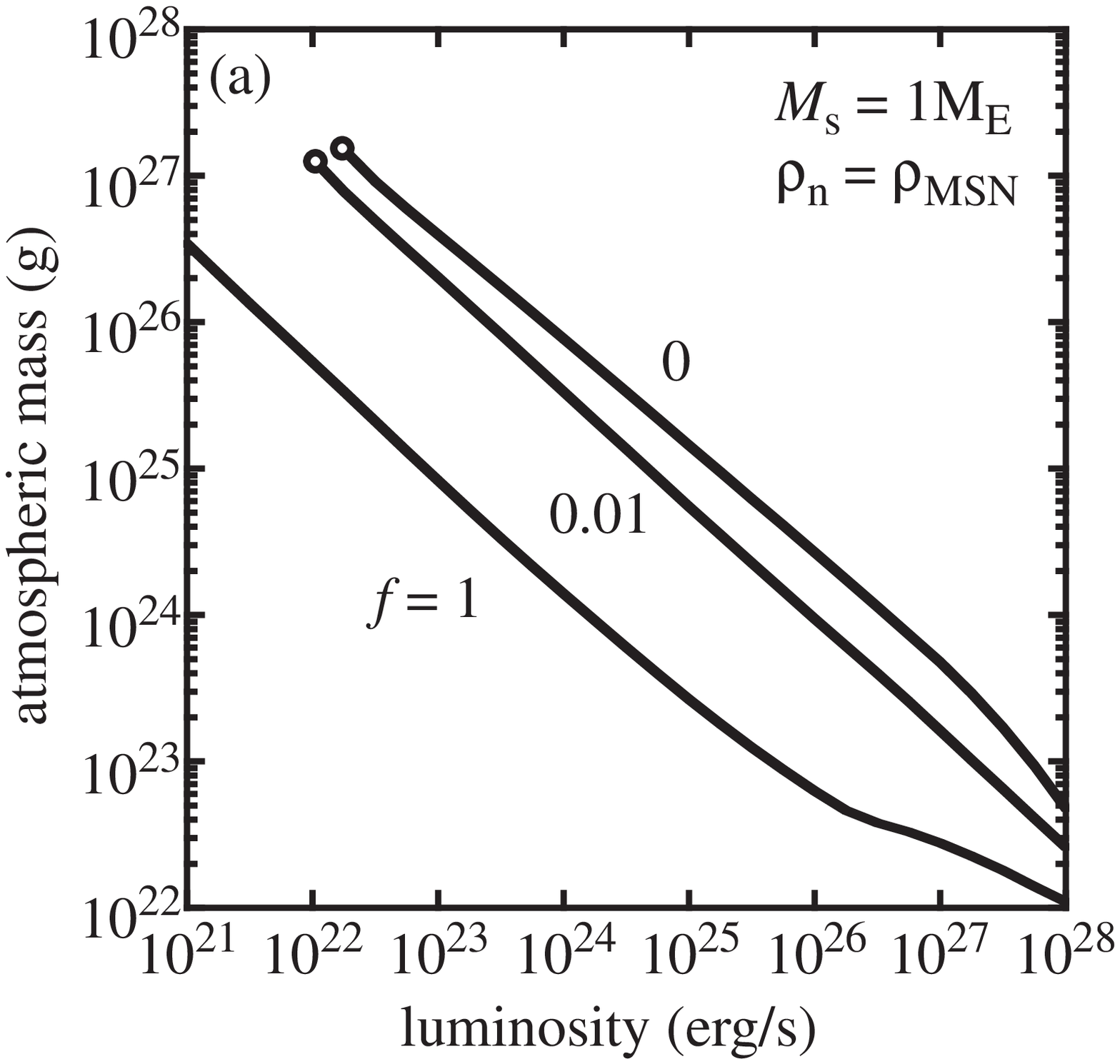}
\includegraphics[width=8cm,keepaspectratio]{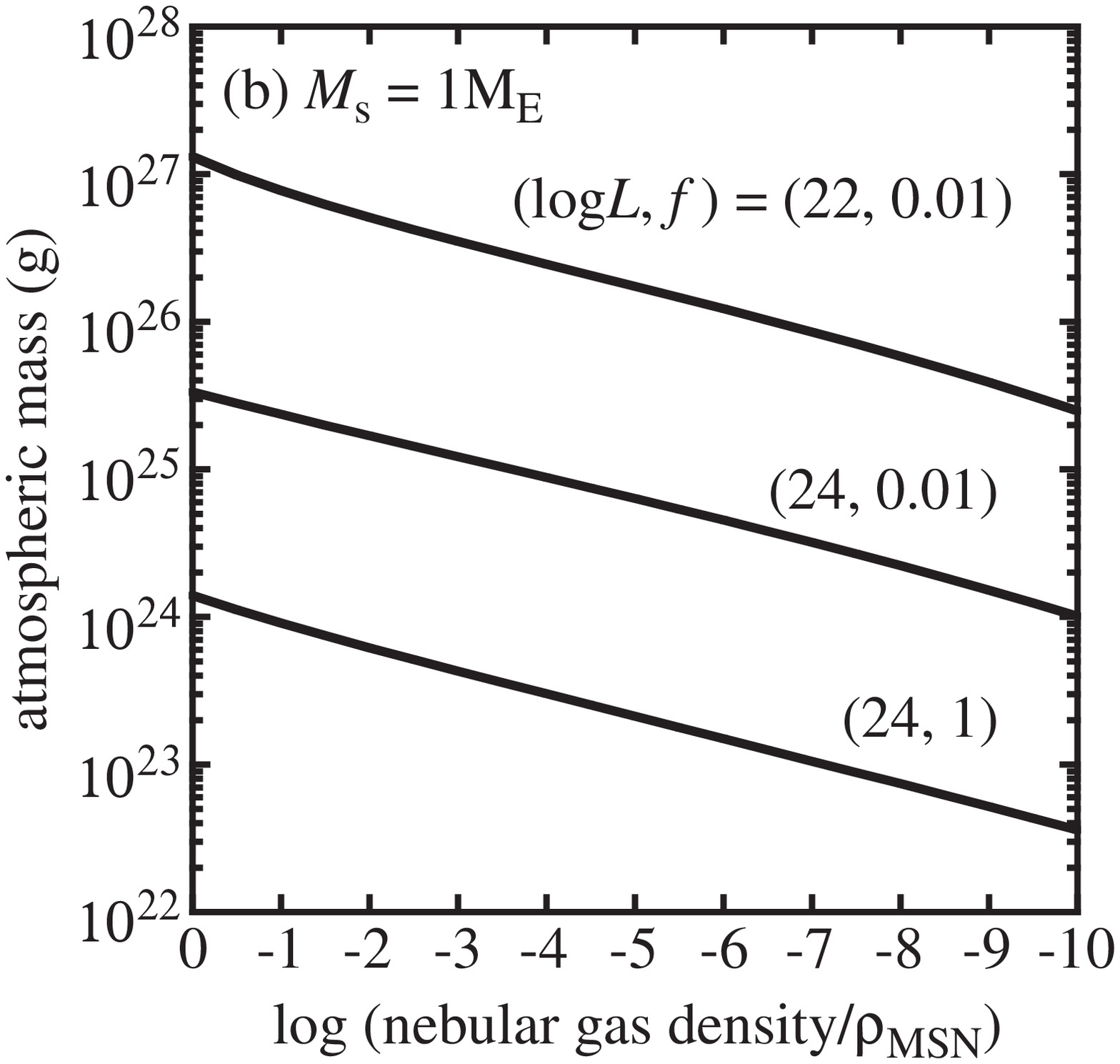}
\caption{The atmospheric mass of an Earth-mass planet 
for wide ranges of three parameters, 
luminosity ($L$), grain depletion factor ($f$), 
and density of the nebular gas ($\rho\sub{n}$). 
In (a), the atmospheric mass is shown as a function of $L$ 
for three different values of $f$. 
In (b), the atmospheric mass is shown 
as a function of $\rho\sub{n}$ 
normalized by that in the minimum-mass solar nebula 
($\rho\sub{MSN} = 1.2 \times 10^{-9} \rm g \, cm^{-3}$) 
for three different sets of values of $L$ and $f$.
The attached sets of values are $\log L$ in $\rm erg\,s^{-1}$ 
and $f$.
Circles represent critical luminosities (see the text). 
} 
\label{f3}
\end{figure}

The surface temperature for $f = 1$ is clearly lower than that for $f \leq 0.01$. 
To understand the difference, we compare the atmospheric structure 
for $f = 1$, 0.01, and $1\times10^{-4}$ in Fig.~\ref{f2}. 
This figure illustrates that the low surface temperature 
for $f = 1$ comes mainly from 
sudden decreases in temperature gradient due to evaporation of silicate 
around 2000~K. 
For $f \leq 0.01$, pressure increases substantially in the optically-thin layer 
and is so high at the evaporation temperature 
that convection takes place at that point; 
the evaporation of silicate thus has no influence on the structure. 

The values of surface temperature obtained here ($<$ ~3300~K) 
are lower than $\sim$ 4000~K obtained by \citet{HNM79}.
Their analytical calculation for a polytropic atmosphere also yielded $\sim$ 4000~K. 
In Fig.~\ref{f2} we show the atmospheric structure 
calculated in the similar way to \citet{HNM79}, 
using \citet{M80}'s gas opacity and 
the grain opacity of $1 \times 10^{-4} \rm cm^2\,g^{-1}$ 
for $T \leq$ 1500~K. 
The grain opacity of $1 \times 10^{-4} \rm cm^2\,g^{-1}$ 
is lower than gas opacity even in the low-temperature region. 
As described in section~2, our gas opacity is much higher than 
that used by \citet{HNM79}. 
Because of the lower opacity in \citet{HNM79}'s model, 
the isothermal layer is deeper, resulting in a large increase in pressure. 
As a result, in \citet{HNM79}'s calculation, 
the atmosphere is almost fully convective except for the isothermal layer. 
As shown in Appendix, surface temperature for a convective atmosphere 
is slightly higher than that for a radiative atmosphere 
(e.g., eq.~[\ref{eq: radiative surface temperature}]). 
This may be the reason why 
the surface temperature for $f = 1 \times 10^{-2}$ and $1 \times 10^{-4}$ 
is slightly lower than that obtained by \citet{HNM79}. 

As shown in Fig.~\ref{f1}b, 
the surface temperature of an Earth-mass planet 
is rather insensitive to the nebular density. 
A decrease in the nebular density certainly lowers 
surface temperature because of a decrease in optical depth, 
but it changes only by a factor of less than two 
even if the nebular density decreases by ten orders of magnitude. 
Figure~\ref{f1}b shows the surface temperature of an Earth-mass planet 
is always higher than 1500~K, the typical melting temperature of silicate, 
for a wide range of the nebular density. 
The insensitivity of the surface temperature of an Earth-mass planet to 
luminosity, opacity, and the nebular density is analytically explained 
in Appendix. 
As also shown in Appendix, it is a rather robust conclusion that 
the surface temperature of an Earth-mass planet is higher than 
the melting temperature of silicate. 

As shown in Fig.~\ref{f3}a, 
atmospheric mass decreases almost linearly with luminosity and 
grain depletion factor (for $f > 0.01$). 
Lower luminosity and opacity yield gentler temperature gradient 
(see eq.~[\ref{eq: radiative temperature gradient}]). 
Then, density gradient must be steeper 
to maintain sufficiently large pressure gradient that supports the gravity. 
Since the density is fixed at the outer edge, 
the steep density profile results in a massive atmosphere. 
It should be noted that in Fig.~\ref{f3}a a curve ends off 
at a critical low value of luminosity at which 
the atmospheric mass is comparable with the planetary mass. 
Beyond the point no static solution is found. 
That means substantial gas accretion takes place 
if luminosity is smaller than the critical value 
(see section~4). 

Atmospheric mass decreases as the nebular density decreases, 
which is shown in Fig.~\ref{f3}b. 
However the dependence is found to be rather weak. 
Most of the atmospheric mass is concentrated in the deep atmosphere. 
And, as shown in Appendix, 
the structure of the deep atmosphere is insensitive 
to the outer boundary conditions. 

\begin{figure}
\includegraphics[width=8cm,keepaspectratio]{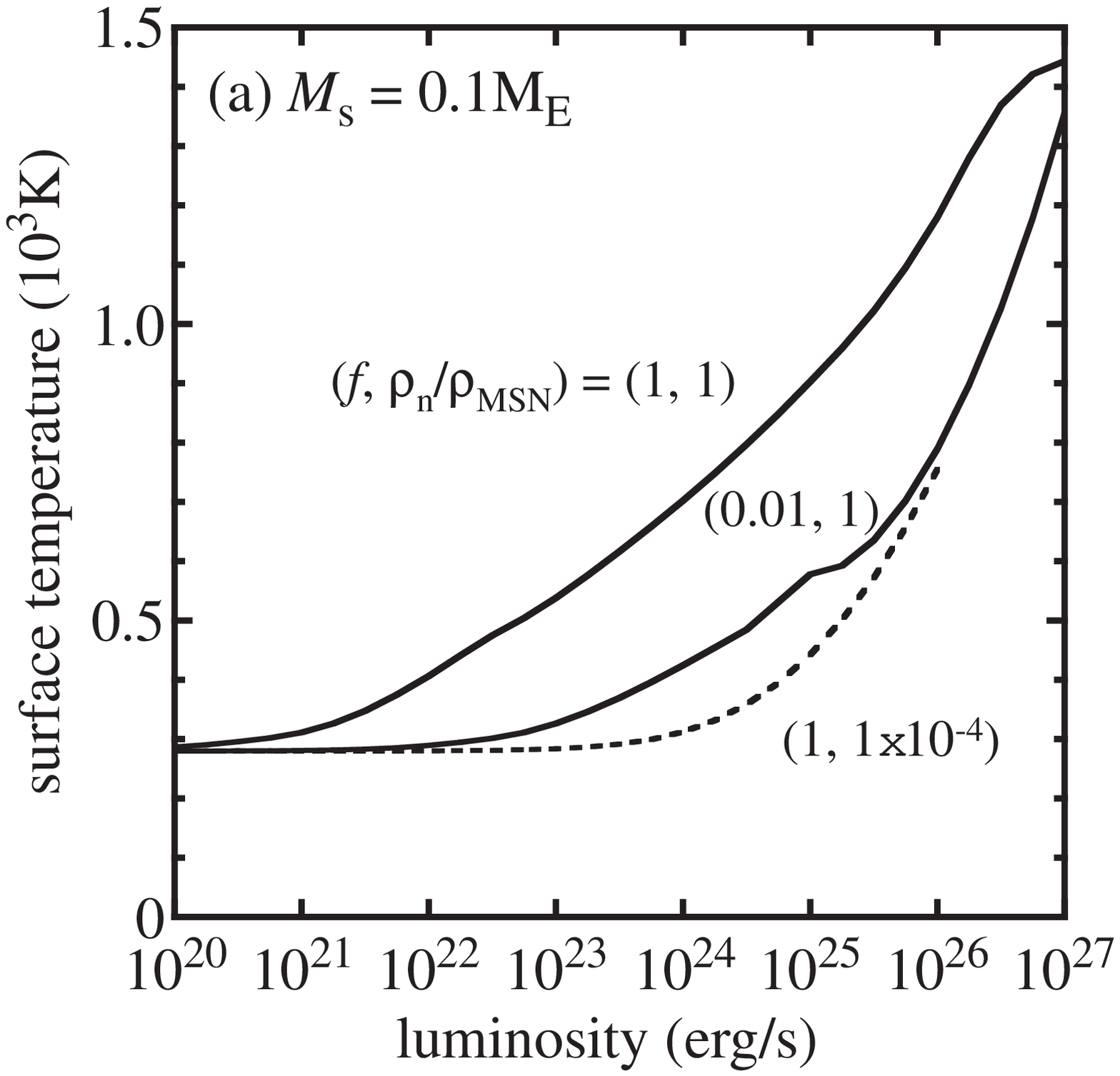}
\includegraphics[width=8cm,keepaspectratio]{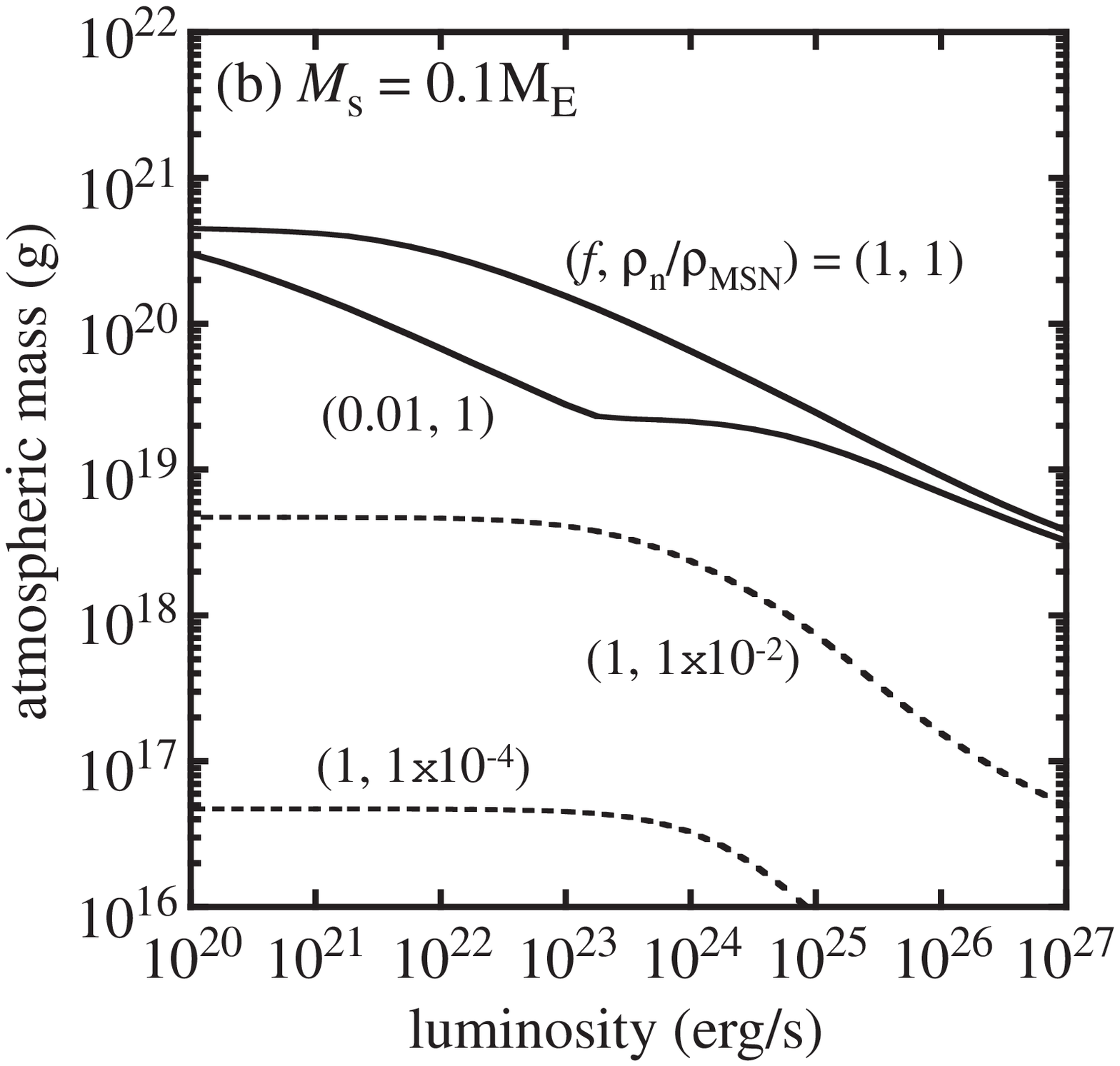}
\caption{The surface temperature (a) and atmospheric mass (b) of 
a 0.1~$\ME$ planet are shown as functions of luminosity for 
several different sets of values of grain depletion factor ($f$) 
and density of the nebular gas ($\rho\sub{n}$) normalized by 
that in the minimum-mass solar nebula 
($\rho\sub{MSN}=1.2\times 10^{-9} \rm g\,cm^{-3}$).
}
\label{f4}
\end{figure}

%\subsection{Case of a Mars-Mass Planet}
The atmosphere on a Mars-mass ($\sim 0.1 \ME$) planet exhibits different behaviour 
compared to that on an Earth-mass planet shown above. 
Figure~\ref{f4} shows surface temperature (a) and 
atmospheric mass (b) as functions of luminosity 
for several choices of grain depletion factor ($f$) and 
the nebular density ($\rho\sub{n}$). 
Unlike the case of an Earth-mass planet, 
the surface temperature of an Mars-mass planet 
is sensitive to luminosity and grain opacity. 
The dependence of the atmospheric mass of an Mars-mass planet 
on those parameters is different from 
that of an Earth-mass planet. 
In Fig.~\ref{f4}a 
surface temperature is shown to appreciably decrease as luminosity decreases. 
Although atmospheric mass increases with decreasing luminosity 
in the similar way to the Earth-mass case for high luminosities, 
there is upper limits to atmospheric mass. 
Comparing Figs.~\ref{f4}a and \ref{f4}b, 
we find that when the atmospheric mass reaches the upper limit, 
surface temperature reaches the nebular temperature ($= 280$~K), 
which means the atmosphere is almost isothermal as a whole. 
As shown in Fig.~\ref{f4}, 
the surface temperature and atmospheric mass of a Mars-mass planet 
are also sensitive to the nebular density, 
which is also different from the Earth-mass case. 
The atmospheric mass decreases almost linearly 
with decreasing nebular density. 
A mathematical explanation for the difference between Earth-mass and Mars-mass 
cases is given in Appendix.

%\subsection{Other Cases}
Figure~\ref{f5} shows surface temperature as a function of luminosity 
for four choices of planet's mass (0.3, 0.5, 0.8, and 1~$\ME$), 
two values of $f$ (0.01 and 1), and $\rho\sub{n} = \rho\sub{MSN}$. 
Smaller planetary mass yields lower surface temperature 
for given $L\sub{s}$ and $f$ 
(e.g., see eq.~[\ref{eq: radiative surface temperature}]). 
The surface temperature for $M\sub{s} = 0.8\ME$ exhibits 
the ``Earth-type'' behaviour that surface temperature is not so sensitive to 
a decrease in luminosity, 
while that for $M\sub{s} = 0.3\ME$ exhibits the ``Mars-type`` behaviour 
that surface temperature is sensitive to a decrease in luminosity. 
The case of a $0.5\ME$ planet seems to be marginal. 
As regards melting of the surface of the solid planet 
that occurs when surface temperature is higher than typically 1500~K, 
the melting is very unlikely on a $0.3\ME$ planet even if 
the planet is embedded in the nebula of density as high as 
that of the minimum-mass solar nebula, 
as shown in Fig.~\ref{f5}. 

%The physical nature of the atmosphere of this kind changes around 0.3-0.5~$\ME$. 
%Figure~\ref{f5} shows atmospheric mass as a function of luminosity (a) 
%and surface temperature as a function of atmospheric mass (b) 
%for several choices of planet's mass. 
%Figure~\ref{f5}a shows there are upper limits to atmospheric mass 
%for planetary mass of $\leq 0.3 \ME$, 
%while curves end at some points for planetary mass of $\geq 0.5 \ME$. 
%Beyond the point, there is no purely-hydrostatic solution. 
%That means that if luminosity is lower than that at the point, 
%contraction of the atmosphere takes place, 
%resulting in the onset of substantial accretion of nebular gas. 
%Figure~\ref{f5}b shows surface temperature is kept to be higher than 
%1500~K on $\geq 0.5 \ME$ planets. 

%
%
%
\section{TIMESCALE FOR GAS ACCRETION} 
As mentioned in section~3, no static solution is found below 
a critical value of the luminosity. 
That means 
energy supply due to atmospheric contraction is needed to 
keep the atmosphere in hydrostatic equilibrium. 
The contraction results in substantial accretion of the nebular gas to 
make the planet a gas-giant one. 
In reality, this happens once the planet becomes isolated from planetesimals. 
In order to exclude gas giant planets from the group of habitable planets, 
we investigate the timescale for the gas accretion 
(i.e., the timescale for a planet to grow up to be a gas giant planet). 
The gas accretion is known to occur not always in a runaway fashion: 
Its typical timescale increases considerably, 
as the mass of the solid planet decreases \citep{INE00}. 

The timescale for the gas accretion also depends on the opacity. 
As described in section~2, 
\citet{INE00} used a rather simple form of the grain opacity 
adopted by \citet{M80}, 
while we use more complicated and realistic grain opacity 
given by \citet{PMC85}. 
The values of \citet{PMC85}'s grain opacity are larger 
by a factor of approximately 10 
than those of \citet{M80}'s for a temperature range 
from $\sim 200$~K to $\sim 1000$~K. 
Also, unlike \citet{M80}'s grain opacity, 
\citet{PMC85}'s depends on gas density 
because the evaporation temperature of rock becomes high 
with gas density. 

Figure~\ref{f6} shows the typical timescale for the gas accretion ($t\sub{g}$) 
after planetesimal accretion is suddenly terminated; 
on this timescale a planet becomes a Jupiter-like planet. 
Following \citet{INE00}, 
we define $t\sub{g}$ as follows. 
We first calculate the purely-hydrostatic structure of the atmosphere 
for given values of $M\sub{s}$ and 
$L\sub{s}$ (see eq.~[\ref{eq: inner boundary conditions}]). 
Then, setting $L\sub{s}$ to be zero, 
we start quasi-static simulation of 
the evolution and accumulation of the atmosphere. 
The luminosity emitted from the outer edge of the atmosphere soon decreases to 
a minimum value and, then, increases progressively 
\citep[see Fig.~2b of][]{INE00}. 
Most of the gas-accretion phase is spent 
when the luminosity is near the minimum value. 
The minimum luminosity is almost equal to 
the above-mentioned critical luminosity \citep{INE00}. 
The characteristic time for growth of atmospheric mass---$M\sub{a}/\dot{M}\sub{a}$ 
where $M\sub{a}$ is atmospheric mass and 
$\dot{M}\sub{a}$ is gas accretion rate---at the minimum luminosity 
obtained by the quasi-static calculations 
are plotted with open circles in Fig.~\ref{f6}. 
%%%
Although we are unable to obtain the timescale for the gas accretion 
for $f < 0.01$ because the simulations for $f < 0.01$ is quite time-consuming, 
the result is expected to be similar to that for $f = 0.01$, 
because difference in $f$ ($< 0.01$) yields only small difference in 
atmospheric mass (see Fig.~\ref{f3}a).
%%%

The timescale $t\sub{g}$ is approximated by \citep{INE00}
\begin{equation}
t\sub{g} = \alpha \frac{G M\sub{s} M\sub{a}^\ast}{R\sub{conv}^\ast L^\ast},
\label{eq: tauG}
\end{equation}
where $L^\ast$ is the critical luminosity, 
$M\sub{a}^\ast$ and $R\sub{conv}^\ast$ are 
the atmospheric mass and outer radius of the inner convective layer 
at $L = L^\ast$. 
The curves in Fig.~\ref{f6} are drawn using equation~(\ref{eq: tauG}) 
with the value of each quantity obtained by our static calculation 
and $\alpha = 1/3$. 
Those curves are fitted roughly by
\begin{equation}
t\sub{g} = 1 \times 10^{10} f \left( \frac{M\sub{s}}{\ME} \right)^{-3.5} 
{\rm yr}
\label{eq: runaway gas accretion}
\end{equation}
for $f \geq 0.01$. 
Not only the value of $t\sub{g}$ itself but also its dependence 
on the mass of the solid planet 
obtained here are different from those by \citet{INE00} 
in which $t\sub{g} \sim 1\times10^8 f (M\sub{s}/\ME)^{-2.5}$~yr. 
In particular $t\sub{g}$ for $M\sub{s} = 1 \ME$ is 100 times 
as large as that given by \citet{INE00}. 
This is because our grain opacity is larger than theirs. 
And this is also because our grain opacity depends on gas density unlike theirs. 
For smaller planetary mass, the critical luminosity is small. 
Then the density for a given temperature is high relative to high-luminosity cases, 
so that the evaporation temperatures are also high. 
Thus the effective grain opacity is higher for smaller planetary masses. 

Similar calculations for a few values of planet's mass 
were also done by \citet*{HBL05}. 
The equation of state and the gas and grain opacity tables they used were 
almost the same as those we have used here. 
We compare our results with 
their results of the duration ($t\sub{g}\super{HBL}$) between 
the end of Phase 1 (at which planetesimal accretion was suddenly terminated) and 
the crossover point (at which envelope mass is equal to core mass) 
in the models named 10H5 and 10H10 in which core masses are approximately 
$5 \ME$ and $10 \ME$, respectively, and $f = 1$. 
Although we are unable to exactly compare our results to theirs 
because the definition of $t\sub{g}\super{HBL}$ is not exactly equal to 
that of $t\sub{g}$, 
both values are found to be similar: 
For $M\sub{s} = 5 \ME$, 
$t\sub{g} \simeq 62 \rm Myr$ while $t\sub{g}\super{HBL} \simeq 78 \rm Myr$; 
for $M\sub{s} = 10 \ME$, 
$t\sub{g} \simeq 4 \rm Myr$ while $t\sub{g}\super{HBL} \simeq 3 \rm Myr$. 

\section{DISCUSSION}

Based on the numerical results obtained in sections~3 and 4, 
we discuss production of water from the hydrogen-rich atmosphere 
on a terrestrial planet, 
a possible range of the mass of a habitable planet, 
and the possibility of the nebular origin of water on the Earth. 

%\subsection{Production of Water on a Earth-Mass Planet}

As described in Introduction, 
production of water on a terrestrial planet requires 
a sufficient amount of hydrogen and 
surface temperature higher than the melting temperature of silicate ($\sim$ 1500~K). 
The two conditions are found to be fulfilled on an Earth-mass planet. 

In a late stage of terrestrial planet formation, 
accretion rate of planetesimals (i.e., luminosity) decreases with time 
because of exhaustion of the planet's feeding zone. 
The decrease in luminosity increases the amount of hydrogen 
on an Earth-mass planet (see Fig.~\ref{f3}a), 
while surface temperature remains above 2000~K (see Fig.~\ref{f1}a). 
For example, if accretion rate of planetesimals is 
$1 \times 10^{-9} \ME \, {\rm yr^{-1}}$, 
corresponding to $L \sim 1 \times 10^{23} \rm erg \, s^{-1}$ 
(see eq.~[\ref{eq: accretion luminosity}]), 
the mass of atmospheric hydrogen is 
more than about $1 \times 10^{25}$~g for $f < 1$, as shown Fig.~\ref{f3}a:  
The amount of hydrogen a planet acquires is insensitive to 
the nebular density (see Fig.~\ref{f3}b). 

Water is produced through reaction between the atmospheric hydrogen 
and oxides contained in the solid planet. 
The amount of water depends on what kind of oxide is available. 
Ion oxides (e.g., w\"{u}stite [$\rm Fe_{0.974}O$], magnetite [$\rm Fe_3O_4$], etc.) 
and fayalite ($\rm Fe_2SiO_4$) 
react with the atmospheric hydrogen to produce water 
comparable in mass to hydrogen; 
the ratios of the partial pressures, 
$P\sub{H_2O}/P\sub{H_2}$, are 0.88, 24.02, and 0.49 at 1500~K for 
the iron-w\"{u}stite ($\rm 1.894Fe+O_2 \leftrightarrow 2Fe_{0.947}O$), 
the w\"{u}stite-magnetite ($\rm 6.696Fe_{0.974}O+O_2 \leftrightarrow 2.174Fe_3O_4$), 
and the quartz-iron-fayalite oxygen 
($\rm 2Fe+SiO_2+O_2 \leftrightarrow Fe_2SiO_4$) buffers, 
respectively \citep*{RHF78}. 
Thus, if a planet acquires hydrogen of $\sim 1 \times 10^{25} \rm g$ and 
such oxygen buffers are available, 
the planet obtains water comparable in mass to 
the current sea water on the Earth ($= 1.4 \times 10^{24}$~g).
However, for the silicon-periclase-forsterite buffer 
($\rm 2MgO+Si+O_2 \leftrightarrow Mg_2SiO_4$), 
$P\sub{H_2O}/P\sub{H_2}$ is as small as $\sim 3 \times 10^{-7}$ \citep{RHF78}. 
Fe-bearing minerals might be required to produce sufficient water, 
although how much water is needed for a planet being habitable is quite uncertain. 
Whether Fe-bearing minerals commonly exist in extrasolar systems is 
still a matter of controversy. 
Equilibrium condensation in a highly-reduced environment 
like a protoplanetary nebula yields not Fe-bearing minerals but Fe-metal 
\citep{WH93}. 
However, dust grains in a protoplanetary nebula can be considered to 
have non-equilibrium composition including, at least, fayalite 
\citep[][and references therein]{PHBSRF94}.

When the surrounding nebular gas disappears almost completely, 
the atmosphere and solid planet begin to get cold, 
and then an ocean forms through the condensation of steam in the atmosphere. 
However, almost complete dissipation of the nebular gas allows 
the extremely ultraviolet (EUV) and far-UV radiation 
from the parent star to penetrate the planetary atmosphere. 
Such irradiation causes extensive loss of hydrogen (and steam). 
The timescale for complete loss of a $10^{25}$-g (say) hydrogen-rich atmosphere 
due to EUV and far-UV can be estimated to be longer than $10^6$~years 
\citep*{SNH80,SHN81}, 
if very strong EUV and far-UV 
\citep[up to 100 times higher than the present;][]{GR02} 
from the young parent star is considered. 
On the other hand, 
the timescale for the ocean formation is on the order of $10^3$~years 
on a planet located in the HZ, 
according to calculations based on the radiative-convective equilibrium model 
of an $\rm H_2O$-$\rm CO_2$ atmosphere with mass of $\sim 10^{24}$~g \citep{A93}. 
The timescale for the ocean formation for a $\rm H_2O$-$\rm H_2$ atmosphere 
considered here is probably not so different from 
that for an $\rm H_2O$-$\rm CO_2$ atmosphere, 
because inclusion of $\rm H_2$ hardly affects the atmospheric structure 
due to its weak blanketing effect. 
Therefore, an ocean can forms before the significant loss of the atmosphere 
due to EUV and far-UV.

We can constrain a possible range of the mass of a habitable planet. 
As shown in section~3, 
there is a lower limit to the planetary mass 
below which the water production proposed in this paper does not work. 
Figure~\ref{f5} illustrates that 
surface temperature of a planet of $\leq$ 0.3~$\ME$ is lower than 
the melting temperature of silicate ($\sim 1500$~K) 
for reasonable ranges of the parameters. 
On the other hand, 
an upper limit to the mass of a habitable planet can be constrained 
because a massive planet captures a huge amount of the nebular gas 
to be a gas giant planet, not a terrestrial planet. 
The timescale for the gas accretion depends strongly on the planetary mass, 
as shown Fig.~\ref{f6}. 
This timescale should be compared to the lifetime of the nebula 
that is known to be about $1 \times 10^7$~years \citep{NGM00}. 
The comparison suggests that the upper limit to the planetary mass is 
$7 \ME$ for $f = 1$ and $2 \ME$ for $f = 0.01$. 

%\subsection{Origin of Water on the Earth}
The water production proposed in this paper may have worked on our Earth. 
Because of \textit{N}-body simulations of planetary accretion, 
details of the terrestrial planet formation in the solar system have been clarified. 
After the runaway growth of protoplanets \citep{WS93}, 
they grow in an oligarchic fashion until they eat almost all of the planetesimals 
in their feeding zones \citep{KI98,KI00}. 
Then several Mars-mass protoplanets form in the terrestrial planet region. 
The subsequent growth of the protoplanets needs giant impacts between them. 
The planets formed in the way is likely to have high eccentricities \citep{CWB96}. 
Damping of those high eccentricities needs the drag force of the nebular gas 
\citep*{KI02,NLT05}. 
Because the Earth is isolated in the nebular gas, 
the accretion of the nebular gas inevitably takes place 
and water is produced on the Earth. 
Although the amount of the nebular gas required 
for the damping of the eccentricities are 
as small as $10^{-4}$ to $10^{-3}$ times that of the minimum-mass solar nebula 
\citep{KI02,NLT05}, 
our numerical results show that this small amount of the nebular gas is sufficient 
for the Earth to get water comparable in mass with Earth's sea water. 

We should be, however, careful when we consider 
the nebular origin of water on the Earth. This is because 
the ratio of deuterium to hydrogen (D/H) of the sea water on the present Earth 
is larger by about a factor of seven than D/H of the solar nebula 
\citep[e.g.,][]{DR02}. 
Moreover, the current Earth's atmosphere includes a tiny amount of noble gas, 
while the solar nebula was rich in noble gas \citep[e.g.,][]{P91}. 
The latter problem may be solved by extensive loss of the noble gases as well as 
hydrogen from the atmosphere 
due to EUV and far-UV radiation \citep{SNH80,SHN81}. 
%At the same time, such escape involves the mass fractionation of the atmosphere 
%\citep[e.g.,][]{Z93}, 
%so that deuterium (D) becomes enriched in the atmosphere. 
%Furthermore, the isotopic exchange between $\rm H_2$ and $\rm H_2O$ 
%followed by the hydrodynamic escape of hydrogen can make the deuterium 
%in the ocean enriched by a factor 2-7 relative to the initial D/H \citep{GI05}. 
%Such deuterium-enrichment in the ocean may solve the former problem, 
%namely the difference in D/H between the present Earth's ocean 
%and the solar nebular gas. 
%Moreover, there is another idea to solve the problem with the difference of D/H. 
Although adequate mixing of water originated from nebular gas 
and water in comets can produce the present D/H on the Earth's ocean, 
because D/H in comets is larger (by about a factor of two) than 
D/H of the present Earth's ocean \citep[e.g.,][]{DR02}, 
the former problem is difficult to solve. 
At present, we have no definite evidences 
that the sea water of the Earth was originated from the nebular gas. 

Our intention in this paper is to claim that 
water production from the nebular gas on a planet  
is a possible way for a terrestrial planet in the HZ to acquire water. 
As shown in this paper, 
the water production works, 
if a planet of 0.3 to several $M\sub{E}$ forms in the protoplanetary nebula. 
The probability of formation of terrestrial planets in the nebular gas 
is still open to debate, 
mainly because the dissipation mechanism of the nebular gas is quite uncertain. 
However, the existence of many gas giant planets in extrasolar planets has 
ensured the validity of the core accretion model. 
That is, accretion of planets generally occurs in a protoplanetary nebula. 
Also, there is no good reason for terrestrial planet formation 
to prefer vacuum environment. 
Therefore, it is rather likely that terrestrial planets also form 
in the surrounding nebular gas. 
The production of water from the nebular gas in the HZ thus seems 
to be a natural consequence of planet formation.

%The probability of delivery of sufficient water strongly depends on 
%the relative position of the snow line to a gas giant planet \citep{RQL04}. 
%Moreover, there is the possibility that the orbital elements of giant gas planets 
%changed drastically after their formations \citep[e.g.,][]{MW02}. 
%Such orbital change makes it more complicated to discuss 
%the probability of water delivery. 
%On the other hand, 


%% Included in this acknowledgments section are examples of the
%% AASTeX hypertext markup commands. Use \url without the optional [HREF]
%% argument when you want to print the url directly in the text. Otherwise,
%% use either \url or \anchor, with the HREF as the first argument and the
%% text to be printed in the second.

\acknowledgments

We are grateful to S. Ida and M. Fujimoto 
for fruitful discussion and their continuous encouragement. 
The manuscript benefited from constructive comments 
from anonymous reviewers. 
We also wish to acknowledge helpful discussion with 
K. Nakazawa, Y. Abe, T. Tanigawa, and H. Senshu. 
This research was partly supported by the 21st Century COE Program 
``How to build habitable planets'', 
Tokyo Institute of Technology, 
by Grand-in-Aid for Scientific Research on Priority Areas, 
both of which are sponsored by the Ministry of Education, Culture, Sports, 
Technology and Science (MEXT), Japan, 
and by the Research Fellowship of the Japan Society for the Promotion 
of Science for Young Scientists.


%% copy editing.  Individual instruments can be provided in parentheses,
%% after the keyword, but they will not be verified.

%Facilities: \facility{Nickel}, \facility{HST(STIS)}, \facility{CXO(ASIS)}.


\appendix
\section{Analytical solutions of the atmospheric structure}

Analytical consideration of the atmospheric structure 
gives a deep insight into the atmospheric properties shown in section~3. 
Here we focus on surface temperature and check whether 
surface temperature exceeds the melting temperature of silicate 
($\sim$ 1500~K). 
In the following analytical consideration, 
we make some additional assumptions: 
(1) The atmospheric self-gravity is negligible; 
(2) the atmospheric gas is perfect gas 
with constant mean molecular weight and adiabatic exponent; 
(3) opacity and luminosity are constant. 
The first assumption is reasonable 
because the atmospheric mass is negligibly small 
relative to the mass of a solid planet 
in cases of our interest. 
Although the second assumption breaks down for $T > 2000$--3000~K 
because of dissociation of $\rm H_2$, 
we adopt it because the temperature range of interest is $<$ 1500~K. 
The third assumption is just for simplicity, but 
the opacity does not change so much 
(within less than one order of magnitude)
for $T < 1500 \rm K$.

We adopt the following dimensionless variables, 
\begin{equation}
\begin{array}{lllcl}
\displaystyle{\varpi = \frac{P}{P_0}}, & 
\displaystyle{\sigma = \frac{\rho}{\rho_0}}, & 
\displaystyle{\theta = \frac{T}{T_0}}, & 
\mbox{and} & 
\displaystyle{x = \frac{r}{R_0}},
\end{array}
\label{eq: nondimensional quantities}
\end{equation}
where quantities with suffices 0 are reference quantities. 
On the above assumptions, using those variables, 
the hydrostatic equation is written by 
\begin{equation}
\frac{1}{\sigma}\frac{d\varpi}{dx} = -\frac{V_0}{x^2},
\label{neq: hydrostatic equilibrium}
\end{equation}
and equation of state is written by 
\begin{equation}
\varpi = \sigma \theta.
\label{neq: equation of state}
\end{equation}
The equations for convective and radiative energy transports  
are respectively given by  
\begin{eqnarray}
\varpi &=& \sigma^\gamma, \label{neq: convective} \\
\frac{d\theta^4}{d\varpi} &=& W_0, \label{neq: radiative}
\end{eqnarray}
In the above set of equations, 
\begin{equation}
V_0 = \frac{GM\sub{s}\rho_0}{R_0P_0} = \frac{GM\sub{s}\mu m\sub{H}}{kR_0T_0},
\end{equation}
where $\mu$ is the mean molecular weight, $m\sub{H}$ is the mass of a hydrogen atom, 
and $k$ is the Boltzmann constant; 
\begin{eqnarray}
W_0 &=& \frac{3\kappa L P_0}{16 \pi \sigma\sub{SB} G M\sub{s} T_0^4} \label{eq: W0}\\
    &=& 4.3 \times 10^{-3}
\left( \frac{M\sub{s}}{1\ME} \right)^{-1}
\left( \frac{\kappa}{1 \rm cm^2\,g^{-1}} \right)
\left( \frac{L}{10^{24} \rm erg\,s^{-1}} \right)
\left( \frac{P_0}{1 \rm Pa} \right)
\left( \frac{T_0}{280 \rm K} \right)^{-4}.
\end{eqnarray}
For later discussion, we define an additional quantity $\lambda$ as
\begin{eqnarray}
\lambda &=& \frac{GM\sub{s}\rho_0}{R\sub{s}P_0} \label{eq: lambda} \\
        &=& \frac{\gamma}{\gamma-1} \frac{R\sub{B}}{R\sub{s}} 
         =  56 \left( \frac{M\sub{s}}{1\ME} \right)^{2/3} 
               \left( \frac{T_0}{280 \rm K} \right)^{-1} 
               \left( \frac{\mu}{2.34} \right) 
               \left( \frac{\rho\sub{s}}{3.9 \rm g\,cm^{-3}} \right)^{1/3},
\end{eqnarray}
where we have used 7/5 for $\gamma$. 

We can obtain the exact solutions for the set of 
equations~(\ref{neq: hydrostatic equilibrium})--(\ref{neq: radiative}). 
The convective solution is given by
\begin{equation}
\theta = 1 + \frac{\gamma - 1}{\gamma} V_0 \left( \frac{1}{x} - 1 \right),
\label{neq: fully convective atmosphere}
\end{equation}
while the radiative solution is given by 
\begin{equation} 
\frac{1}{x} = 1 + \frac{1}{V_0} 
\left\{ 4 \left(\theta -1 \right) + F\left(\theta, W_0 \right)\right\},
\label{neq: fully radiative atmosphere}
\end{equation}
where 
\begin{equation}
F (\theta, W_0) = 
w_0 \left[
\ln \left( \frac{\theta-w_0}{\theta+w_0} \frac{1+w_0}{1-w_0} \right) 
-2 \left( \arctan \frac{\theta}{w_0} - \arctan \frac{1}{w_0} \right) 
\right] 
\end{equation}
for $W_0 < 1$; 
\begin{equation}
w_0 \equiv \left( 1 - W_0 \right)^{1/4}.
\end{equation}
The exact solution for $W_0 > 1$ can be also obtained \citep{II03}.

The dimensionless surface temperature ($\theta\sub{s}$) 
for a fully convective atmosphere is given by 
\begin{equation}
\theta\sub{s} = \frac{1}{x\sub{s}} = \frac{\gamma-1}{\gamma} \lambda,
\label{neq: convective approximate solution}
\end{equation}
since $R_0 = R\sub{B}$ and thus $V_0 = \gamma/(\gamma-1)$; 
$\theta\sub{s}$ for a fully radiative atmosphere is given by 
\begin{equation}
\theta\sub{s} = \frac{V_0}{4} \frac{1}{x\sub{s}} 
              = \frac{\gamma}{4 (\gamma-1)} \frac{1}{x\sub{s}} 
              = \frac{\lambda}{4},
\label{neq: radiative approximate solution}
\end{equation}
if $\theta\sub{s} \gg 1$ and $x\sub{s} \ll 1$. 
The dimensionless surface temperature is found to be determined 
only by $\lambda$. 
Thus, surface temperature is given by
%\begin{equation}
%\theta\sub{s} = \left\{ 
%\begin{array}{l}
%0.29 \\ 0.25
%\end{array}
%\right\} \lambda, 
%\end{equation}
%that is, 
\begin{equation}
T\sub{s} = \left\{
\begin{array}{l}
4500 \\ 3900 
\end{array} \right\}
\left( \frac{M\sub{s}}{1\ME} \right)^{2/3}
\left( \frac{\mu}{2.34} \right) 
\left( \frac{\rho\sub{s}}{3.9 \rm g\,cm^{-3}} \right)^{1/3} \rm K,
\label{eq: radiative surface temperature}
\end{equation}
where the upper and lower values are for the convective ($\gamma = 7/5$) 
and radiative cases, respectively. 
Those solutions were also given by \citet{HNM79}. 
The surface temperature given by equation~(\ref{eq: radiative surface temperature}) 
is determined only by planet's mass 
(for given $\mu$ and $\rho\sub{s}$) and 
independent of luminosity, opacity, and outer boundary conditions. 

The above conclusion is consistent with the numerical results 
for an Earth-mass planet 
but in contradiction with those for a Mars-mass planet 
given in section~3. 
To understand the discrepancy, 
we compare the exact solution (\ref{neq: fully radiative atmosphere}) 
with the approximate one (\ref{neq: radiative approximate solution}); 
the result is shown in Fig.~\ref{f7}. 
The difference is found to depend on the value of $W_0$. 
For an Earth-mass planet, the difference is only 40~\% 
even for $\log W_0 = -10$ and 
the exact value of $T\sub{s} \sim$ 2300~K, 
while, for a Mars-mass planet, 
the difference is more sensitive to $W_0$ and 
the exact $T\sub{s}$ reaches $T_0$ at $\log W_0 = -4$. 
From a mathematical point of view, 
the sensitivity is more remarkable for smaller value of $\lambda$. 

The existence of the outermost isothermal layer lowers surface temperature. 
To check how much surface temperature is reduced, 
we simulate the structure of a two-layer atmosphere composed of 
the outer isothermal layer and the inner radiative or convective layer. 
We first determine the radius of the photosphere 
where the optical depth defined by equation~(\ref{eq: optical depth}) is 
equal to 2/3, using the isothermal solution 
\begin{equation}
\sigma = \exp \left[ V_0 \left( \frac{1}{x}-1 \right) \right]. 
\end{equation} 
Once we obtained values of the quantities at the photosphere, 
checking convective stability, 
we calculate surface temperature 
using equation~(\ref{neq: fully convective atmosphere}) or 
(\ref{neq: fully radiative atmosphere}). 
Figure~\ref{f8} shows the surface temperature 
as a function of $\rho\sub{n}/\rho\sub{MSN}$ 
for six different values of $\kappa$ and $L = 1\times10^{21} \rm erg\,s^{-1}$; 
The solid and dashed lines representing 
convective and radiative lower layers, respectively. 
If we adopt higher values of luminosity, 
we obtain higher temperatures for $\kappa \ge 1 \times 10^{-4} \rm cm^2\,g^{-1}$. 
As shown in Fig.~\ref{f8}, 
even if luminosity is as small as $1 \times 10^{21} \rm erg\,s^{-1}$, 
surface temperature is above the melting temperature of silicate 
($\simeq$ 1500~K) except for a limited range of the nebular density, 
$\rho\sub{n} < 1 \times 10^{-6} \rho\sub{MSN}$.




\clearpage

\begin{figure}
\includegraphics[]{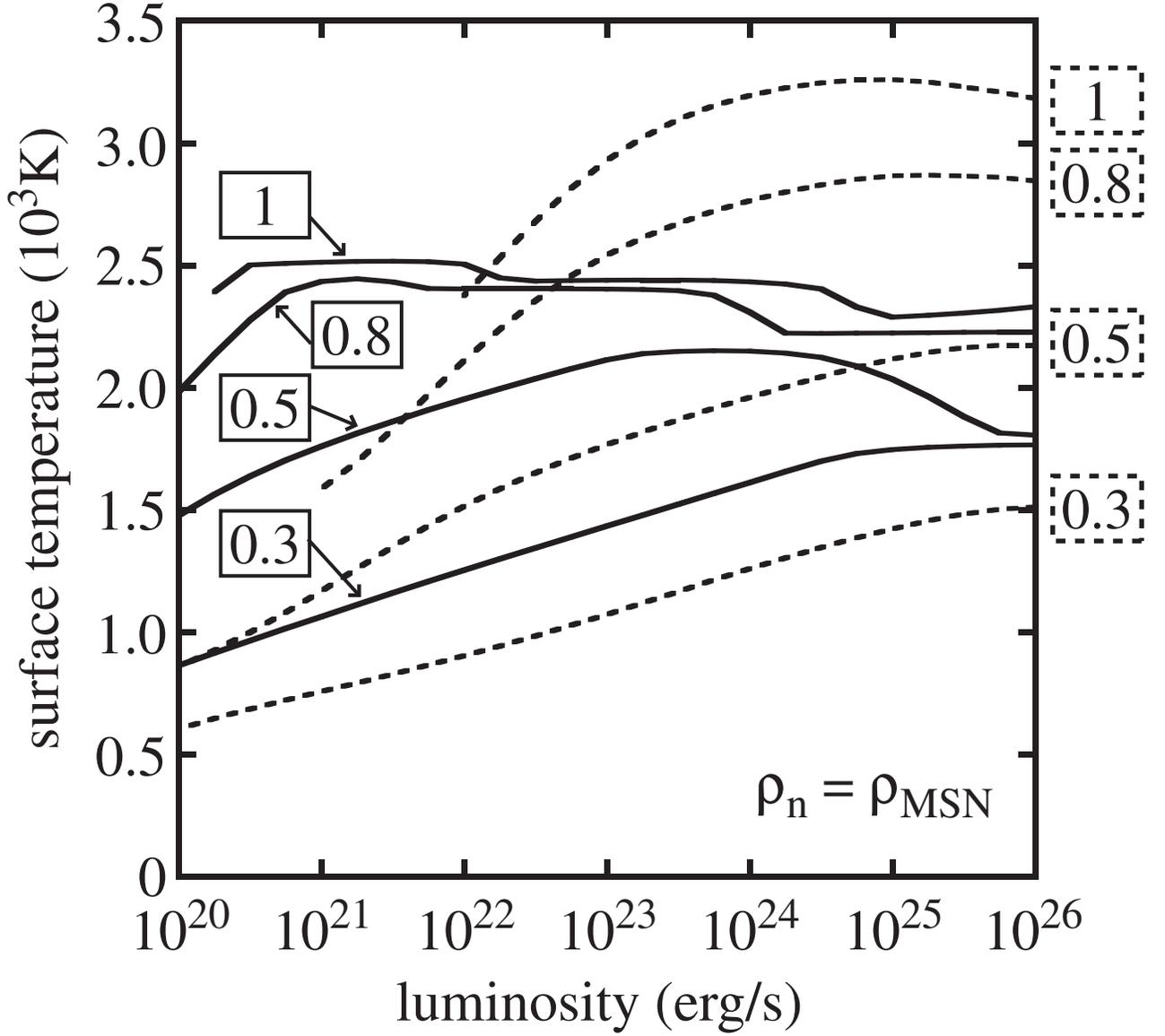}
\caption{
Surface temperature as a function of luminosity 
for several choices of planet's mass. 
Each attached number represents planet's mass in the Earth mass. 
The solid and dashed lines represent 
cases of $f = 1$ and $0.01$, respectively. 
}
\label{f5}
\end{figure}

\begin{figure}
\includegraphics[]{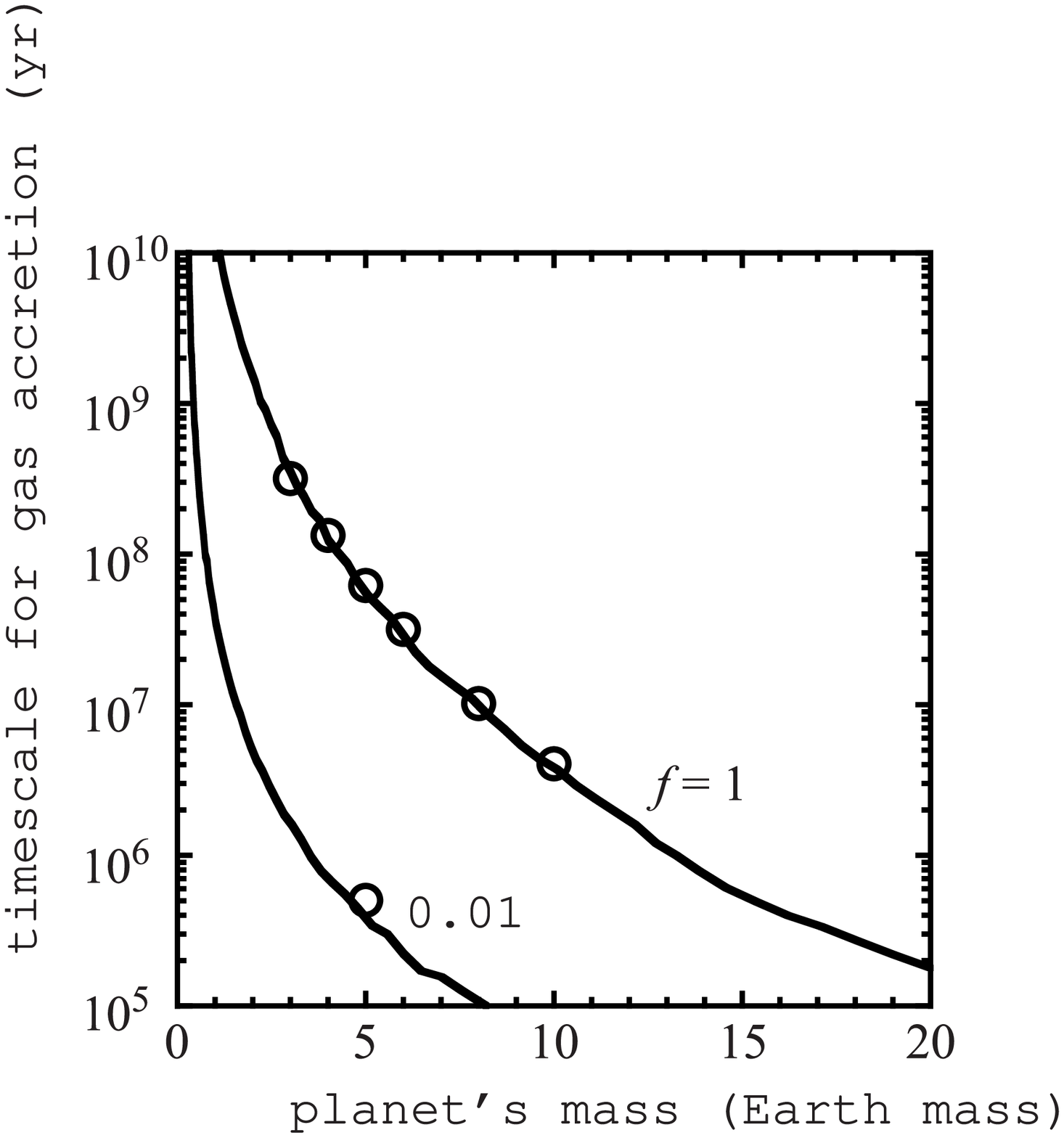}
\caption{The typical timescale for substantial accretion of the nebular gas 
as a function of planet's mass for two choices of the grain depletion factor 
($f$). 
The open circles represent the numerical results of our evolutionary calculations; 
the solid lines are drawn using equation~(\ref{eq: tauG}) 
with the numerical factor $\alpha$ of 1/3.}
\label{f6}
\end{figure}

\begin{figure}
\includegraphics[]{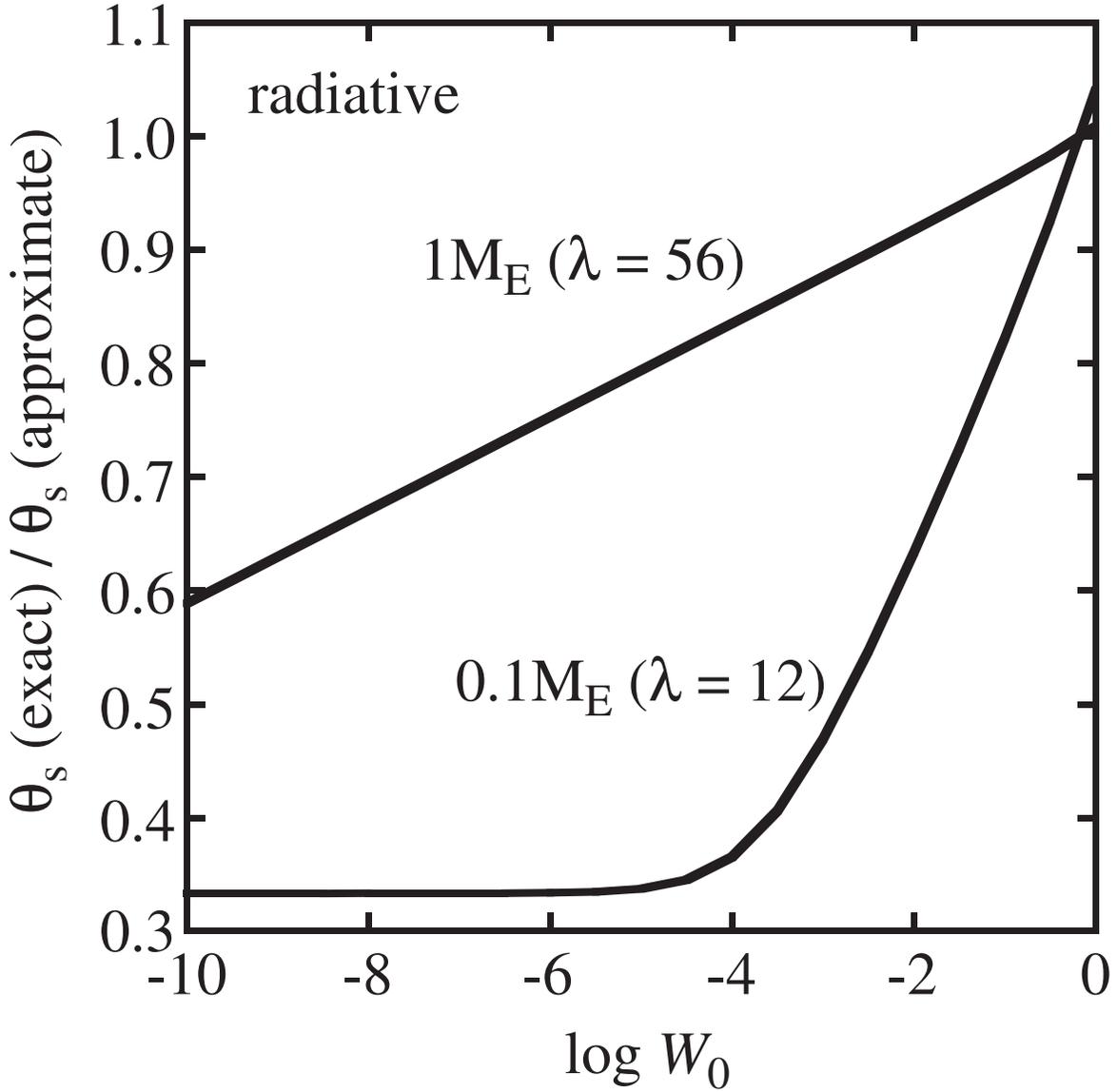}
\caption{Difference in value of the dimensionless surface temperature 
($\theta\sub{s}$) between the exact radiative solution 
(eq.~[\ref{neq: fully radiative atmosphere}]) and 
the approximate radiative solution called the radiative-zero solution 
(eq.~[\ref{neq: radiative approximate solution}]) 
for two cases of 1~$\ME$ and 0.1~$\ME$ planets. 
The definitions of $W_0$ and $\lambda$ are given by 
equations~(\ref{eq: W0}) and (\ref{eq: lambda}), respectively.}
\label{f7}
\end{figure}

\begin{figure}
\includegraphics[]{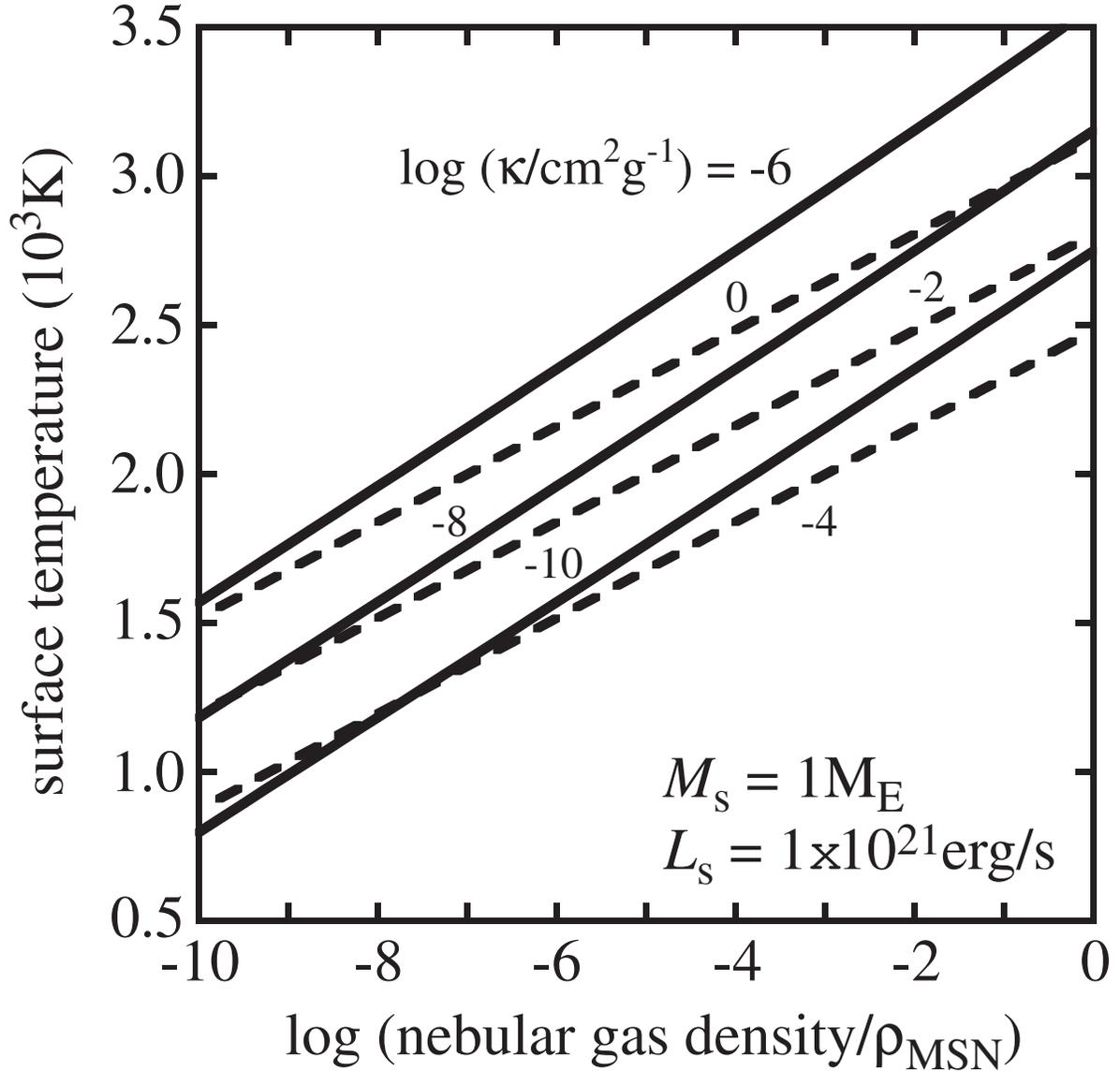}
\caption{The surface temperature of an Earth-mass planet 
as a function of density of the nebular gas 
that is normalized by that of the minimum-mass solar nebula 
($\rho\sub{MSN} = 1.2 \times 10^{-9} \rm g \, cm^{-3}$) 
for six different values of opacity ($\kappa$) and 
luminosity ($L\sub{s}$) of $1\times10^{21} \rm erg\,s^{-1}$ 
that has been analytically obtained 
in the two-layer (outer isothermal and inner radiative or convective) 
atmospheric model (see Appendix). 
The solid and dashed lines respectively 
convective anr radiative inner atmospheres. 
}
\label{f8}
\end{figure}

\end{document}